\documentclass[acmsmall]{acmart}
\AtBeginDocument{%
  \providecommand\BibTeX{{%
    \normalfont B\kern-0.5em{\scshape i\kern-0.25em b}\kern-0.8em\TeX}}}

\setcopyright{acmcopyright}
\copyrightyear{2019}
\acmYear{2019}
\acmDOI{10.1145/xxxxx.xxxxx}
\acmJournal{JACM}
\acmVolume{1}
\acmNumber{1}
\acmArticle{1}
\acmMonth{1}

\usepackage{makecell}
\usepackage{textcomp}
\usepackage[titlenumbered,ruled,noend]{algorithm2e}
\usepackage{mathtools}
\usepackage{bbold}
\usepackage{float}
\usepackage{tikz}
\usepackage{graphicx}
\usepackage{csquotes}
\usepackage{graphics}
\usepackage[english]{babel}
\usepackage{multicol}
\usepackage{multirow}
\usepackage{booktabs}
\usepackage{url}

\begin{document}
\title{A Small Survey on Event Detection Using Twitter}
\author{Debanjan Datta}
\email{ddatta@vt.edu}
\affiliation{%
  \institution{Virginia Tech}
  \city{Arlington}
  \state{VA}
  \country{USA}
}

\begin{abstract}

\end{abstract}

\keywords{Social Media, Data Mining, Microblog}

\maketitle

\section{Introduction}{
The proliferation of social media has affected how people interact, opine, consume and propagate information. This is evident from popular phenomena such as effects of fake news and online social movements. However the the data obtained from social media presents itself with large volume and velocity, accompanied by significant amount of irrelevant data pertaining to general discussions, personal messages and spam. Social media has been shown to be effective for detecting, forecasting and tracking real world events. The ability to detect real world events is crucial and has applications in disease surveillance, commerce, governance and other areas. Thus extraction of useful information and modelling the characteristics of social media to detect real world events is an important problem.
}
\section{Research Problem}{
To outline the research problem we need to define events, which has multiple interpretations. We define event as a real world occurrence characterized by time, context and possible location. Events can be categorized as open domain or targeted domain, such as health or politics\cite{ritter2012open}. Events can also be categorized as global events such as online movements, or local events that have a spatial component. Event detection has been studied both in online(detection) and offline(extraction) systems, and are application specific sharing similar techniques. In this work we refer to them as event detection for brevity.
\\
The research problem discussed is detection of unspecified spatio-temporal events from microblogs. The methods discussed use textual content, temporal and spatial features in microblogs, without external data sources such as news. The problem is challenging because text data in social streams is colloquial, with non-standard acronyms, words outside vocabulary and dynamic nature. Additionally there are incorrect capitalization, emoticons and features such as mentions and hashtags. The text is short and unstructured due to limits on text length mandated by platform and due to changing patterns of interpersonal communication. Thus entity resolution is difficult\cite{yamada2015enhancing} compared to structured news data. Individual user inputs do not concisely capture the entire context or event, and often contain event related or non-related subjective sentiments and personal messages. Events evolve in their occurrence period, and text streams reflect this in a near real time, yet in a nonuniform manner. Location of origin for the text and user locations are sparse \cite{compton2014geotagging} as well and event popularity is skewed. The approaches in event detection are based on the understanding that a real world event leads to a significant increase of topically coherent inputs from users which are temporally and spatially coincident with the event. The determination of signals in text stream that signify an event require triangulation of the time, context and location simultaneously make it a challenging problem. We consider methods that have been demonstrated with data microblog sites such as Twitter and Weibo, which have significant adoption rate. There are multiple related problems to event detection such as event theme tracking, first story detection and targeted domain event detection. We do not include methods for specified events, such as disease or political protests that use techniques such as query expansion.
}
\section{Background}{
Microblog stream is a set of temporally ordered documents (${T_1,T_2 ... T_k}$) where k is number of time points and documents have unique discrete timestamps. It can contain text, URLs, hashtags, references to other users along with time and location information. Here we assume that the text is not multilingual. Text data is preprocessed by tokenization and removing non-essential attributes. Location information is present as latitude and longitude. A \textit{gazetteer} or geolocation service is used to obtain the data on the geographic region of focus. Mentions are references by author of one document to another user of the microblog. Hashtags are a word or phrase preceded by a hash sign (\#) relating to a context, topic or entity. Events have a start and end time, and the duration is assumed be co-incident with the signal from the microblog. The methods for event detection utilize techniques from graphs, Bayesian models, Named Entity Recognition(NER)\cite{ritter2011named},\cite{yamada2015enhancing} and time series to name a few.
}
\section{Global Event Detection in Microblogs}{
Event detection has been studied in Topic Detection and Tracking program, where the primary data source was news stories. Some the earlier works that define the problem are presented in \cite{allan1998topic} and \cite{yang1998study}. Multiple models have been proposed on detecting global events in open domain which use techniques such as document clustering, phrase networks and wavelet transformations, but without the spatial aspect. The ideas presented in these prior methods are relevant to the approaches to problem being discussed. One of the earlier works that discusses online event detection is \cite{sakaki2010earthquake}. It uses spatial, temporal and keyword features combined with a Kalman Filter or Particle Filter combined with support vector machine classifier to detect natural disasters. The authors in \cite{aggarwal2012event} present a detailed work on mining events in social streams using text, temporal distribution and network information. However the model does not take in to account spatial information and merges similar events, though spatially not co-located. In \cite{petrovic2010streaming} the model uses uses locality sensitive hashing for fist story detection. In \cite{becker2011beyond} an online clustering algorithm is presented using temporal features, word counts, hash tags to detect events. \textit{PhraseNet}\cite{melvin2017event} present an architecture that follows a similar paradigm, using phrases instead of unigrams derived from microblog documents. \textit{EDCoW}\cite{weng2011event}and \cite{cordeiro2012twitter} present wavelet based approaches. Some of the recent works in this area are \cite{wurzer2015twitter},\cite{nguyen2015real},\cite{ge2016event} and \cite{repp2018extracting}. 
}
\section{Spatio-Temporal Event Detection in Microblogs}{
Event detection methods in microblogs clustering on document, keywords, temporal or spatial features. The majority of methods discussed employ clustering and burst detection. There are two major approaches to find the spatial component of an event. One is where the method first detects an event using document related features, then location is estimated. The other is where location specific documents are first determined, and then temporal and document features are used for event extraction. However in some cases such as Bayesian models, the temporal, semantic and spatial components of documents are simultaneously modelled. The different methods also vary in terms of their geographic granularity, as in some of the models are able to detect events at city level and others at region level. These have been summarized in Table 1. 
\begin{table}
\begin{center}
\begin{tabular}{|l|c|} 

\textbf{Granularity} &\textbf{Models} \\
\toprule
City Level & \makecell{
EvenTweet\cite{abdelhaq2013eventweet}, Streamcube\cite{feng2015streamcube}, \cite{dong2015multiscale},\\ GeoBurst\cite{zhang2016geoburst},\cite{wei2017finding}, TrioVecEvent\cite{zhang2017triovecevent}, GeoBurst+\cite{zhang2018geoburst+} }
\\
\midrule
Region Level & 
\makecell{
TwitterStand\cite{sankaranarayanan2009twitterstand}, \cite{lee2010measuring}, \cite{gao2013novel},\\ LSED\cite{unankard2013location}, \cite{edouard2017graph}, LEM\cite{zhou2014simple}, GTSS\cite{liu2016graph}, DPEMM\cite{zhou2017event}}
\\
\bottomrule
\end{tabular}
\end{center}
\label{table1}
\caption{Spatial Granularity of Event Detection}
\end{table}
\subsection{Document and Feature Clustering} {
Many of the earlier models utilize an online document clustering method, which is suitable because it is unsupervised and no training data is required. The general approach is to cluster temporally and semantically similar documents using a sliding window for extracting documents, using thresholds on semantic and temporal similarity. Semantic similarity is calculated using a similarity measure such as cosine function on shallow features such as vector space representation of documents. These approaches have some common drawbacks as well. Fragmentation occurs in online clustering when documents which should be clustered together are not, and this needs to be handled which has computational overhead. Also, using centroid of term vectors may cluster together similar but separate events when they share significant number of terms in the same time period. The approaches are sensitive to thresholds used for clustering and number of clusters.

\textit{TwitterStand}\cite{sankaranarayanan2009twitterstand} uses online document based clustering algorithm. A Naive Bayes classifier is used to remove irrelevant documents, and includes seed users who publish news. Similarity measure for clustering is cosine similarity between TF-IDF of the document centroid and candidate document, weighted by a Gaussian attenuator on time centroid of the cluster. The model tackles fragmentation using a periodic update to connect similar clusters. Location of event clusters are obtained from documents using geo-tags and toponym resolution, as well as location of authors of documents present in the cluster. Apart from the common drawbacks of document clustering approach, sparsity of user location and geo-tags~\cite{compton2014geotagging} may also affect results. 

\textit{EvenTweet}~\cite{abdelhaq2013eventweet} takes a spatio-temporal keyword clustering based approach for real-time event detection using geo-tagged documents. The model selects bursty terms in a time period. Then their spatial distributions are calculated using a spatial grid and terms with entropy below a threshold are selected, since terms with higher entropy imply more uniform spatial distribution. A hierarchical clustering algorithm (BIRCH\cite{zhang1997birch}) is used to cluster selected terms, using cosine similarity on spatial distribution of each word. The algorithm is sensitive to the value of this entropy threshold. The clustering process allows for efficient computation, however the process requires the number of clusters which is a disadvantage. After clusters are obtained they are ranked using a score. The clusters are scored based on burstiness and spatial coverage of the terms. The top-k clusters are interpreted as event clusters. The creation of spatial grids and calculation of keyword entropy has computational overhead, but it can be employed in online systems. The method clusters keywords without taking into account their semantic coherence and does not utilize hashtags, which may affect results.

In \cite{gao2013novel} the model uses a spatio-temporal document clustering approach using adaptive K-means\cite{bhatia2004adaptive}. The model first calculates normalized geographical burstiness of documents, and selects areas with burstiness above a threshold. Then the documents are clustered using cosine similarity of TF-IDF vectors. Clusters with document counts over a user defined threshold are interpreted as events. Although the number of clusters is not required, the algorithm uses multiple constants that may not generalize well. The clustering algorithm requires pairwise similarity calculation of all documents to set the minimum distance threshold for assignment of documents to clusters, which is computationally expensive. Also subtle events are ignored if the threshold for burstiness is not set appropriately.

\textit{LSED}\cite{unankard2013location} presents an online event detection method similar to ones discussed above. NER is first used to extract document location, and hierarchical agglomerative clustering is used to cluster documents based on semantic similarity which precludes the need to specify number of clusters. The selection of event clusters from candidate clusters takes a different approach. A correlation score is calculated between event location and the static location of authors in the cluster obtained from user profiles, which takes into account the hierarchical structure of locations(regions,city,etc). Cluster correlation scores are calculated using these, and top $k$ clusters are selected. Among these, the clusters with temporal burstiness is terms of number of documents are selected as emerging events. The algorithm may not be able to detect events where users are mobile e.g. protest where people congregate, or there is sparse user location information. 

\textit{StreamCube}\cite{feng2015streamcube} presents an online multi-scale event detection approach using hashtags clustering in geo-tagged documents. It considers time, latitude and longitude as the three dimensions of a cube, where both time and space have a hierarchical structure. The geographical region is divided into grids and implemented using a quad tree like structure. A sliding window is used to extract documents, placed at finest granularity and clustered using hashtags. The similarity metric for clustering is a weighted average of hashtag co-occurrences and word co-occurrences associated with hashtags in the documents. Events are considered as a cluster of hashtags, and represented as a set of words and hashtags. The clustering algorithm utilizes the hierarchical structure by placing documents at the lowest level first and then merging clusters to create clusters of lower granularity. Event clusters are merged based on nearest neighbor approach, with a predefined threshold. Each event is then ranked, using a linear combination of document count, burstiness and spatial concentration. The output is the top-k events in a cube. The method provides an efficient approach for dynamic clustering hashtags and words given evolving content, and merging of events. The assumption made that hashtags are sufficient to capture an event may lead to incorrect inference where volume or co-occurrence of hashtags are not significant. The model does not consider the semantic coherence between related documents unless they share hashtags, and is sensitive to the weights that are user defined. 

Social media users are important sources for detecting local events, and this is used in \cite{wei2017finding}. The model first calculates the location of users using the method presented in \cite{compton2014geotagging}. It analyzes the the text stream to find presence and weight of reciprocal relationships between authors and authors with geo-tagged documents. Filtering out noise, the static locations of authors are calculated as L1-median of the geo-tags of their documents, and these are the labelled users. Then a spatial label propagation algorithm is used, that assigns a static location label to each unlabelled user in the network based on multivariate L1-median of its neighbors within a threshold of maximum distance. It alleviates the issue sparsity of geo-tagged documents using author locations. In the event detection stage, the documents are assigned locations if not geo-tagged. Then an online clustering algorithm employs a temporally weighted TF-IDF vector similar to \textit{TwitterStand}\cite{sankaranarayanan2009twitterstand} and detects events. This method is suitable for comparatively small geographical areas of interest with a large number of users. The method however does not account for new users being added over time, or users being inactive or mobile.

\textit{GeoBurst}\cite{zhang2016geoburst} and \textit{GeoBurst+}\cite{zhang2018geoburst+} use geo-topic clusters obtained from geo-tagged documents to detect events in an online system. The models use a query window to extract documents, and each document is considered a tuple of location, time and keywords. Geo-topic clusters, which are candidate events, are created using the semantic and spatial coherence between documents. Spatial coherence between documents is obtained using an Epanechnikov kernel on the geo-locations. Semantic coherence is obtained using efficient implementation of node specific Pagerank\cite{lofgren2013personalized} on the keyword co-occurence graph. Documents with highest combined score in the neighborhood are termed as pivot and first local pivots for each document are calculated. Then global pivot for each document is calculated using an iterative algorithm. The pivots and their neighboring documents are considered as clusters, since they are semantically and spatially coherent. An activity timeline is constructed that maintains the candidate clusters, and handles fragmentation and temporal validity of clusters. Cluster features such as temporal and spatial mean and deviation, and keyword occurrences are extracted. In \textit{GeoBurst}, a ranking function based on cluster features is used to extract top $k$ clusters which are interpreted as events. \textit{GeoBurst+}\cite{zhang2018geoburst+} uses a different approach. An embedding learner is used to capture document semantics by jointly mapping the document and keywords into the same low-dimensional space, using the concept of \textit{doc2vec}\cite{le2014distributed}. These are combined with the cluster features to obtain spatial and temporal burstiness, keywords and spatio-temporal anomaly. These are fed into a linear regression classifier to obtain event clusters. Crowd-sourced annotated data is used to train a linear regression classifier to classify events. The works presents efficient methods which are scalable to online systems and extracts a multitude of useful features that provide better results than competing approaches such as \cite{abdelhaq2013eventweet} and \cite{dong2015multiscale}on demonstrated datasets.
}
\subsection{Bayesian Mixture Models}{
\textit{Latent Event Model(LEM)}\cite{zhou2014simple} the authors propose a generative model with latent variable to detect events. The latent variable is the event membership of individual documents. The model first preprocesses document using stemming, NER and part of speech tagging to extract entities (location and non location) and keywords. Each event is considered as tuple of items - entity($y$), location($l$), keywords ($k$) and time($d)$. For each document, an event $e$ is drawn from the multinomial distribution of events $\pi$, which is obtained from a Dirichlet prior. For each event $e$, a corresponding set of multinomial distributions ($\theta_e$,$\phi_e$,$\psi_e$,$\omega_e$) are drawn from respective Dirichlet priors, such that $y$ $\sim$ $\theta_e$, $d$ $\sim$ $\phi_e$, $l$ $\sim$ $\psi_e$ and $k$ $\sim$ $\omega_e$. Each document is modeled as a joint distribution over $y$, $d$, $l$ and $k$. Gibbs sampling\cite{hoff2009first} is used for estimating the parameters. The output is the clusters of documents. In the post processing step an entity, date, location, and top two keywords with highest probability are extracted as events. 
\\
  \textit{Dirichlet Process Event Mixture Model(DPEMM)}\cite{zhou2017event} is built upon \textit{LEM}\cite{zhou2014simple}. Each event is considered as tuple of items - entity($y$), location($l$), keywords ($k$) and time($d)$. A generative model is proposed, which uses the same notation and distributions as in \textit{LEM}. The key difference between this model and \textit{LEM} is that generation of a new event follows the Chinese Restaurant Process in this case, rather than being selected only from a prespecified set of events thus eliminating need to specify number of events. A second model \textit{DPEMM-WE} model is also proposed where locations and entities are jointly represented using reduced dimension embedding. Their respective distributions are assumed to be Gaussian in the generative process, since related items should be closer in embedding space. Both the models estimate the parameters using Gibbs sampling. The output of clustering process is a cluster of documents. Post processing is applied to the clusters, and events extracted using the co-occurrence of the items with highest probability.
\\
The above three models are unsupervised, hence annotated data is not needed to train the model. However they rely on entity recognition and part of speech tagging. Moreover, the assumption that $y$, $l$ and $k$ are mutually independent may not hold in certain cases. They also do not utilize features such as hashtags and author relations. Moreover, they have not been demonstrated to be implemented in an online system.
\\
  \textit{TrioEvecEvent}\cite{zhang2017triovecevent} presents a online local event on geo-tagged documents detection method using multi-modal embedding and Bayesian mixture model. It uses multimodal embedding and classifier, similar to \cite{zhang2018geoburst+}. The embedding jointly maps all the spatial, temporal, and textual elements present in a document into the same low-dimensional space with their correlations preserved. Then a Bayesian mixture model divides the documents in query window into $K$ geo-topic clusters which correspond to candidate events. In the generative model, cluster membership($z_d$) for documents are drawn from multinomial distribution $\pi$, which has a Dirichlet prior. Locations are drawn from a Gaussian distribution for that event. Each document is modelled as tuple of keywords and location. Semantic embedding for keywords are drawn from vonMisesFisher distribution for the event, which is effective for representing directional data in the vector space. The parameters are obtained using Gibbs Sampling. After obtaining the candidate clusters using the posterior distributions for $z_d$, a logistic regression classifier is used to separate the event clusters from non-event clusters. It uses features such as spatial and temporal surprisngness, bursty nature and semantic coherence of the clusters. The algorithm employs an incremental approach while shifting query window for recalculation of clusters, which is suitable for online applications. Although the number of clusters is an input, the authors note that some of the clusters may be empty thus addressing the issue by overestimation of $k$. Annotated data is required for training the classifier, and this is obtained through crowd-sourcing. The authors demonstrate model's performance to be better than \textit{EvenTweet}\cite{abdelhaq2013eventweet}, \cite{dong2015multiscale}, \textit{GeoBurst}\cite{zhang2016geoburst} and \textit{GeoBurst+}\cite{zhang2018geoburst+} on two large datasets. \\
\textit{TweetSCAN}\cite{capdevila2017tweet} combines a generative model with DBSCAN to detect events. However the work does not provide any comparative or quantitative evaluation. There are a few other works which follow approaches very similar to works discussed in this section \cite{guo2016nonparametric},\cite{yilmaz2018multimodal}. 
}
\subsection{Graph Based Models}{
  Two online event detection methods are proposed in \cite{dong2015multiscale}, where event detection is modelled as a graph-based clustering problem. The first model is aimed at local event detection. A graph ($G$) is constructed using documents as nodes, and edge weights as cosine similarity between the documents using TF-IDF vectors. The documents are selected within a small time window and spatial boundary. $G$ is clustered using Louvain clustering\cite{blondel2008fast} to optimize local modularity, with the assumption that events are localized in both time and space. 
The second model presented utilizes concept of discrete wavelet transform to detect events at different spatial and temporal scales. This is different from the multi-resolution model in \cite{feng2015streamcube}. The key idea here is that if two documents are semantically similar, a trade-off is tolerable in computation of either the spatial or temporal proximity in terms of resolution. The geographical region is divided into multi-resolution grids, and predefined number of spatial and temporal scales are chosen. The spatial scale is determined by the spatial distance between two cells, and a finer temporal scale leads to coarser spatial scale and vice-versa. Terms are extracted from the documents, and the time series of their counts are calculated at the finest spatial and temporal scales. Haar transform\cite{graps1995introduction} is applied to each time series, to obtain coefficients at multiple granularities which enable efficient time series similarity calculation. The spatial similarity between documents is calculated as maximum of similarities between shared keywords. Thus the similarity between documents is obtained as product of temporal, spatial and semantic similarity, and act as edge weights in the graph. Louvain clustering\cite{blondel2008fast} is applied to obtain clusters of documents, which are then processed to remove noise such as frequent non-event words. Each obtained cluster corresponds to an event. The calculation of wavelet transform and time series similarity in keywords has computational overhead. The method, as shown by subsequent works is suitable for detecting localized events with a longer time span.
\\
  \textit{Graph Topic Scan Statistic(GTSS)}\cite{liu2016graph} combines spatial scan statistics\cite{wang2008spatial}, topic modelling and graph clustering. The key idea is that the topical distribution of words in an area during a event would be different from the background topical distribution. The locations are formulated as a graph, with nodes as location(cities) and neighboring locations connected by an edge. The geo-tagged documents are assigned to respective nodes. The topic model used is mixture of unigrams\cite{hofmann1999probabilistic} model where each document belongs to a single topic, and the background distribution is calculated from historical data. A generative model is used where documents from event area and non event area have different topic distributions, the distribution of words, documents and topics are jointly modelled for each location. The problem is formulated as finding the sub-graph(cluster) that maximizes the spatial scan statistics, and solved using an iterative approximation procedure. The spatial scan statistic is defined as $log(\frac{argmax_{\theta}P(Data|H_1(S),\theta)}{P(Data|H_0,\pi)})$. $H_1$ is the hypothesis that a region S has topic distribution $\theta$ and $H_0$ is the hypothesis that all the regions have the same topic distribution $\pi$. Documents are geo-tagged using NER and author profiles, which addresses the sparsity of geo-tagged documents. However the work does not demonstrate ability to find multiple events within same time window in a connected region or multiple types of events in an online system. Another disadvantage is that data over a long time period is required to train the model.
\\
  In \cite{edouard2017graph} a graph clustering based approach for event detection is proposed. The model preprocesses documents with NER, spelling corrections and entity linking. The entities include location, persons and organizations. Keywords surrounding named entities are selected using a window. A directed graph is created where nodes are entities and selected keywords, with edge weights as co-occurrence counts. Edges also keep a list of reference to relevant documents. Graph partition is performed on $G$, and strongly connected sub-graphs obtained with the assumption that sub-graphs are not related. A modified Pagerank\cite{page1999pagerank} algorithm is used to assign scores to nodes. The nodes with scores above a threshold are considered as candidate events, and sequentially processed from highest score. For each candidate node, its successor and predecessor with maximum scores are selected. Using these and documents (stored in edges), an event is defined. The location, entities and time are extracted using NER and timestamps from the event set. The model's performance is demonstrated to be favorable compared against \textit{LEM}\cite{zhou2014simple} and \textit{DPEMM}\cite{zhou2017event}. The method does not demonstrate applicability in an online system, and features such as hashtags and geo-tags are not utilized. 
\\
}
  Apart from the categories of approaches listed above, an earlier work was \cite{lee2010measuring} which uses geographical patterns of users to detect events. Instead of using a graph or grid of locations, it divides the entire geographical into regions of interests (Voronoi cells) using K-means clustering on geo-locations of users. It then uses region specific tweet counts, user movements and user activity compared with long term behavior to determine events. The method is applicable where user locations are actively observable, and there is active change in those such as geo-social events(e.g. protest). Most events do not necessarily result in or display user movements. Semantic content or hashtags are not used, thus detection of multiple co-occurring events of different types is not supported. The method also employs a set of rules which may be affected by changing user set and user behaviors. There are some other works\cite{mathioudakis2010twittermonitor},\cite{watanabe2011jasmine},\cite{walther2013geo} which use approaches discussed above and focuses on visual analysis, and we exclude them due to space constraints .
}
\section{Model Evaluation}{
Analysis and evaluation is required to understand how the effective implemented concepts in the different steps are and compare the performance against state of the art methods. Measuring the quantitative and qualitative performance of event detection is an area of research\cite{weiler2015evaluation} itself. Though most of the works discussed provide either qualitative or quantitative evaluation, only few provide detailed quantitative and qualitative evaluation. Evaluation results are dependant on the granularity of spatial and temporal scales of the events a model is optimized for, e.g city level events within last hour or regional event within past day. Thus for comparative analysis baselines should be chosen such that they are appropriate for the spatio-temporal scale of the proposed model. Another factor that affects performance of the discussed models are their parameters such as thresholds, which need to be correctly set. Performance analysis on a proposed model is essential if the model has such parameters. For evaluation, ground truth is required which is difficult to obtain and many of the works use manually annotated data sets for this purpose \cite{dong2015multiscale},\cite{feng2015streamcube},\cite{liu2016graph}. Crowd sourcing is more efficient and robust approach given large amounts of data, and this has been used in \cite{zhang2017triovecevent},\cite{zhang2018geoburst+}. Since an event is correctly detected if its time, location and context are correctly detected, the tolerance for error in terms of time and location will define the performance.
\paragraph{Qualitative Evaluation}{
Event detection systems discussed do not always provide a summary, but most significant keywords, hashtags or entities along with time and location. Thus the interpretation of detected events is key part of the evaluation process. Microblog data is used to detect a small number of significant events, and matched against ground truth data. This provides an insight how well the relevant entities, keywords, hashtags apart from time and location that are related to the event are captured. It can also help understand the discriminative capacity of the model for similar co-occurring events. Some of the works such as \cite{feng2015streamcube},\cite{dong2015multiscale},\cite{zhang2017triovecevent},\cite{zhang2018geoburst+} follow this approach. However qualitative analysis of small number of events may ignore subtle events if the model fails to detect them, which may be of interest.
}
\paragraph{Quantitative Evaluation}{
Most of the approaches in literature employ information retrieval based model evaluation metrics precision, recall and F-1 score \cite{zhou2014simple},\cite{dong2015multiscale},\cite{zhou2017event},\cite{zhang2017triovecevent},\cite{edouard2017graph} ,\cite{zhang2016geoburst}. Precision ($P$) is $N_{t}/N_{r}$, where $N_{t}$ is the number of detected events which are true and $N_{r}$ is the total number of events detected. Since it is difficult to obtain complete set of events, pseudo recall($R$) is more appropriate where an ensemble of baseline approaches can be used to obtain a more complete set($N_{e}$), as used in \cite{zhang2017triovecevent},\cite{zhang2018geoburst+}. Pseudo recall is defined as $R = N_{t}/ N_{e}$. Pseudo F-1 score is $((2.P.R)/(P+R))$. Most literature presents F-measure with $\beta =1$(F1-score)\cite{sasaki2007truth}. However for open domain events, where recall is more important $\beta>1$ should be preferred. It should be noted that model should be evaluated against a significant amount of data, and test all spatial and temporal scales that the model is designed to detect events in to ascertain the generalizability of the results. The baselines should also be run with the same inputs. However metrics based on event count may not provide a complete picture where events are merged or duplicate events detected. Computational efficiency should be empirically evaluated against competing baselines, and whether the baselines are online systems or not needs to be accounted for. Methods which select ranking in selecting clusters such as \cite{abdelhaq2013eventweet},\cite{feng2015streamcube},\cite{zhang2016geoburst} require evaluation of ranking performance and Mean Average Precision is an appropriate measure. Clustering performance for documents or hashtags are important in analysis of the model. Metrics like Normalized Mutual Information or Purity\cite{zhou2017event},\cite{dong2015multiscale} are applicable, but require annotation of individual documents which is costly. 
}

}
\section{Conclusion}{
There has been growing interest in event detection due to practical applications. In this work, we discuss the current approaches for spatio-temporal event detection using only microblogs as data source. Since the last part of the decade, there has been a exponential growth in social media use that has led microblogs to be a vital data source for event detection. The nature of microblogs is also evolving with time, for instance Twitter has increased character limit and mobile devices are the most popular hardware platform. The state of the art provides recall and precision of approximately $60\%$ and $80\%$ respectively\cite{zhang2017triovecevent}, which is impressive given the challenging nature of the problem. We demonstrated how different related research areas are relevant to this problem, and this guides future work. There has been growing interest in natural language processing related to microblog data which is relevant to the problem\cite{xia2015discriminative},\cite{zhang2018adaptive}. There has been ongoing research on using multimedia content, multiple data sources to find better results. Representation learning on graphs, rumor detection and multimodal techniques may be explored in future as well. 

}
\bibliographystyle{ACM-Reference-Format}
\bibliography{main}
\end{document}